\documentclass[letterpaper,twocolumn,10pt]{article}
\usepackage{usenix-2020-09}

\usepackage{tikz}
\usepackage{amsmath}

\usepackage{cite}
\usepackage{amsmath,amssymb,amsfonts}
\usepackage{algorithmic}
\usepackage{graphicx}
\usepackage{balance}
\usepackage{textcomp}
\usepackage{xcolor}

\usepackage{tabularx}

\usepackage{amsmath,amssymb,amsfonts}
\usepackage{algorithmic}
\usepackage[ruled,vlined,linesnumbered]{algorithm2e}
\usepackage{graphicx}
\usepackage{textcomp}
\usepackage{xcolor}
\usepackage{url}
\usepackage{array}
\newcolumntype{M}{>{\centering\arraybackslash}m{5.0cm}}
\newcolumntype{N}{>{\centering\arraybackslash}m{3.0cm}}
\newcolumntype{J}{>{\centering\arraybackslash}m{2.0cm}}
\newcolumntype{K}{>{\centering\arraybackslash}m{1.5cm}}
\newcolumntype{L}{>{\centering\arraybackslash}m{1.0cm}}
\usepackage{multirow}
\usepackage{amssymb}
\usepackage[normalem]{ulem}

\usepackage{newtxtt}
\usepackage{color}
\definecolor{deepblue}{rgb}{0,0,0.5}
\definecolor{deepred}{rgb}{0.6,0,0}
\definecolor{deepgreen}{rgb}{0,0.5,0}
\definecolor{lineNumberColor}{rgb}{0.47,0.47,0.47}

\usepackage{xcolor}
 
\definecolor{codegreen}{rgb}{0,0.6,0}
\definecolor{codegray}{rgb}{0.5,0.5,0.5}
\definecolor{codepurple}{rgb}{0.58,0,0.82}
\definecolor{backcolour}{rgb}{0.95,0.95,0.92}

\usepackage{comment}
\usepackage{multirow}
\usepackage{listings}
\usepackage[shortlabels]{enumitem}
\usepackage{subcaption}
\usepackage{ifthen}
\usepackage[most]{tcolorbox}
\usepackage{alltt}
\usepackage{boxedminipage}
\usepackage{tikz}
\usepackage{balance}

\usepackage{authblk}

\usepackage{pgfplots}
\pgfplotsset{compat=1.16}

\DeclareRobustCommand\circleone{\tikz[baseline=(char.base)]{
            \node[shape=circle,draw,inner sep=2pt] (char) {1};}}

\DeclareRobustCommand\circletwo{\tikz[baseline=(char.base)]{
            \node[shape=circle,draw,inner sep=2pt] (char) {2};}}

\newcommand*\circled[1]{\tikz[baseline=(char.base)]{
            \node[shape=circle,draw,inner sep=2pt] (char) {#1};}}
            
\usepackage{amsthm}
\newtheorem{definition}{Definition}

\SetKw{Continue}{continue}
\SetKw{Return}{return}
\SetKw{Function}{function}

\makeatletter
\newcommand{\nosemic}{\renewcommand{\@endalgocfline}{\relax}}
\newcommand{\dosemic}{\renewcommand{\@endalgocfline}{\algocf@endline}}
\let\oldnl\nl
\newcommand{\nonl}{\renewcommand{\nl}{\let\nl\oldnl}}
\makeatother

\usepackage{ifthen}
\definecolor{red}{HTML}{9B0000}
\definecolor{blue}{HTML}{000000}

\newboolean{showcomments}
\setboolean{showcomments}{true}
\ifthenelse{\boolean{showcomments}}
{\newcommand{\mynote}[2]{\textcolor{red}{
			\fbox{\bfseries\sffamily\scriptsize#1}
			{\small$\blacktriangleright$\textsf{\emph{#2}}$\blacktriangleleft$}}}}
{\newcommand{\mynote}[2]{}}

\def\BibTeX{{\rm B\kern-.05em{\sc i\kern-.025em b}\kern-.08em
    T\kern-.1667em\lower.7ex\hbox{E}\kern-.125emX}}

\newcommand{\tool}{\textsc{SkipFuzz}}

\begin{document}

\title{SkipFuzz: Active Learning-based Input Selection for \\ Fuzzing Deep Learning Libraries
}

\author[1]{\rm Hong Jin Kang}
\author[1]{\rm Pattarakrit Rattanukul}
\author[1]{\rm Stefanus Agus Haryono}
\author[1]{\rm Truong Giang Nguyen}
\author[2]{\rm Chaiyong Ragkhitwetsagul}
\author[3]{\rm Corina Pasareanu}
\author[1]{\rm David Lo}
\affil[1]{Singapore Management University}
\affil[2]{Mahidol University}
\affil[3]{Carnegie Mellon University and NASA Ames Research Center}

\maketitle

\begin{abstract}
    Many modern software systems are enabled by deep learning libraries such as TensorFlow and PyTorch.
    As  deep learning is now prevalent, the security of deep learning libraries is a key concern.
    Fuzzing is a promising direction to find library vulnerabilities.
    Fuzzing deep learning libraries presents two challenges.
    Firstly, to reach the functionality of the libraries, fuzzers have to use inputs from the valid input domain of each API function, which may be unknown.
    Secondly, many inputs are redundant as they trigger the same behaviors.
    Randomly sampled invalid inputs are likely not to trigger new behaviors.
    While existing approaches partially address the first challenge, they overlook the second challenge.
    
    We propose \tool{}, a new approach for fuzzing deep learning libraries.
    To generate semantically-valid inputs, 
    \tool{} learns the input constraints of each API function using active learning.
    By using information gained during fuzzing, \tool{} is able to infer a model of the input constraints,
    and, thus, generate valid inputs. 
    \tool{} comprises an active learner which queries a test executor to obtain feedback used to infer the input constraints.
    After constructing hypotheses of the actual input constraints,
    the active learner poses queries and refines the hypotheses using the feedback from the test executor, which indicates if the library accepts or rejects an input, i.e., if it satisfies the input constraints or not.
    Inputs from different \textit{categories} are used to invoke the library to 
    check if a set of inputs satisfies a function's input constraints.
    Inputs in one category are distinguished from other categories by possible input constraints they would satisfy, e.g. they are tensors of a certain shape.
    As such, \tool{} is able to refine its hypothesis by eliminating possible candidates of the input constraints.
    This active learning-based approach addresses the challenge of redundant inputs. 
    To infer the input constraints, the active learner poses only queries that may provide new information for refinement.
    
    Our experiments indicate that \tool{} generates more crashing inputs than previously proposed approaches.
    Using \tool{}, we have found and reported 43 crashes. 28 of them have been confirmed, with 13 unique CVEs assigned.

\end{abstract}

\section{Introduction}

The use of deep learning is now prevalent. It affects many aspects of our lives, including in safety and security-critical domains such as  self-driving vehicles~\cite{sato2021dirty,zhou2020deepbillboard}. 
Consequently, there have been increasing concerns about vulnerabilities in  deep learning systems, 
which can have a severe impact. 
While many studies have focused on testing and uncovering weaknesses of deep learning \textit{models},
the development of approaches that mitigate the risks of vulnerabilities in deep learning \textit{libraries}, 
such as TensorFlow and PyTorch, is equally crucial.
These vulnerabilities may lead to errors that corrupt memory contents or crash the software system, which can be abused for denial-of-service attacks on applications using deep learning libraries.

\textbf{Challenges.}
Fuzzing the deep learning libraries poses challenges  related to the selection of suitable inputs.
The first challenge is that the space of inputs is large and many generated inputs do not belong to the domain of \textbf{semantically-valid inputs}.
By randomly selecting inputs from the large input space, the vast majority of inputs would be  rejected by the library's input validation checks, and therefore, fail to adequately test the library's behaviors.
A second challenge is related to the \textbf{redundancy of inputs}. 
Given the observation of a test outcome (e.g., an exception thrown when invoked with a particular input), an appropriate  strategy should be employed to pick an input that tests a different behavior from the already observed test outcome.
Otherwise, the same library behavior would be tested again, leading to redundancies in fuzzing.
Ideally, a fuzzer triggers a wide range of test behaviors.

For most of the deep learning libraries' APIs, the input constraints are  unknown~\cite{xie2022docter}.
Without knowledge of the input constraints, randomly generated inputs can be supplied. 
However, unlike other programs e.g. UNIX utilities~\cite{miller1990empirical,miller2020relevance} that take sequences of bytes as input,
libraries accept inputs that are highly structured~\cite{padhye2019semantic,babic2019fudge}. 
Likewise, for TensorFlow and PyTorch, randomly generated inputs are unlikely to be structurally valid (e.g. a tensor) or semantically valid (i.e., passing the libraries' input validation).

\textbf{Existing approaches.} Existing works propose  methods of partially addressing the first challenge of selecting inputs satisfying the function's input constraints, i.e., generating valid inputs.
As off-the-shelf fuzzers do not encode knowledge of these constraints and cannot generate a high proportion of semantically valid inputs, 
Xie et al.~\cite{xie2022docter} proposed \texttt{DocTer}, which infers the input constraints from API documentation.
\texttt{FreeFuzz}~\cite{wei2022free} mines valid inputs of functions from open source code and resources.
\texttt{DeepRel}~\cite{deng2022fuzzing}, building on \texttt{FreeFuzz}, 
identifies pairs of similar functions using their documentation to share the mined valid inputs.
\texttt{DocTer} and \texttt{DeepRel} rely on API documentation, which may not always be available or well-maintained.
Therefore, 
they may not be able to cover all  functions in the libraries' APIs~\cite{xie2022docter}.
Moreover, not every function would be invoked in open-source code.
This motivates new techniques of \textit{input constraint} inference  {\em sans} documentation and high-quality sample usages.

Existing approaches do not address the second  challenge of high input redundancy.
They apply random mutations to change the type, value, etc.~\cite{wei2022free,deng2022fuzzing} of a valid input or  randomly select inputs based on the input constraints~\cite{xie2022docter}.
There are numerous possible inputs, with the  majority of them triggering the same behaviors and provides no new information.
This motivates methods of distinguishing inputs and systematically selecting them for invoking the library.

\textbf{Our approach.} In this study, 
we propose an approach (embodied in a tool), \tool{}, that infers a model of the input constraints at the same time as fuzzing the deep learning library.
\tool{} does not require existing specifications or the collection of a wide range of seed inputs, as was done in prior work. 
Instead, it learns the input constraints of the API functions through fuzzing.
Once inferred, the input constraints allow the generation of valid inputs.
To do so, \tool{} employs \textit{active learning}, which learns by interactively querying an oracle.
In \tool{}, the test executor takes the role of the oracle by constructing and executing test cases invoking the library based on the queries.
The \textit{test outcomes} (e.g., if the input is valid, invalid, or a crashing input) are provided back to the active learner.
For successful inference of the input constraints, the active learner 
queries the test executor with a wide range of inputs that satisfy/violate different possible input constraints.
This enables fuzzing with less redundancy.

\tool{} leverages findings 
from prior studies~\cite{jia2020empirical,jia2021symptoms,islam2019comprehensive,xie2022docter}.
\tool{} employs a set of input \textit{properties}, by which test inputs are distinguished. 
The 
design of the input properties is based on the input constraints and root causes of bugs  identified in these studies.
\tool{} seeks to reduce redundancy by assuming that inputs with the same properties trigger the same behaviors; if the input does (not) trigger a vulnerability, then other inputs with the same properties will also (not) trigger it.
The input properties differentiate inputs by  their structure, shapes,  values,
corresponding to possible input constraints.
Inputs satisfying the same  properties are grouped into the same \textit{category}.
Selecting inputs from different categories allows for more input diversity and
is more likely to provide new information about the input constraints.

The active learner aims to identify a hypothesis of the input constraints that is \textit{consistent} with the observed outcomes. 
In the active learning literature~\cite{cambronero2019active}, a consistent hypothesis is one where the behavior of the program  under the hypothesis  matches that of the actual program.
Given the execution history indicating if each input was valid or invalid,
an ideal hypothesis is a set of categories that contain the valid inputs but exclude the invalid inputs.
We quantitatively assess the consistency between  a given hypothesis and the observed test outcomes using {\em precision}, the proportion of observed valid inputs under the hypothesis,  and {\em recall}, the proportion of valid inputs under the hypothesis out of all observed valid inputs.
Once a hypothesis is found to be adequate, \tool{} generates only inputs satisfying the hypothesized input constraint, allowing for a high proportion of valid inputs to be generated.

In our experiments, \tool{} detects crashes in 108 functions in TensorFlow and 58 functions in PyTorch. 
After analyzing and removing crashes with similar root causes, the new crashes have been reported to the developers. 
23 TensorFlow vulnerabilities and 6 PyTorch bug reports have been confirmed or fixed.
\tool{} can trigger up to 65\% of the crashes found by the prior approaches, \texttt{DocTer} and \texttt{DeepRel}.
In a deeper analysis, we find that \tool{} has a greater input and output diversity, which contributes to its capability in generating crashing inputs.
\tool{} is able to generate valid inputs for 37\% of TensorFlow and PyTorch's API, while prior approaches only generate valid inputs for up to 30\% of the API.
When the active learner succeeds in inferring an input constraint,  \tool{} is able to generate valid inputs over 70\% of the time, indicating that the input constraint was inferred reasonably well.
This validates that active learning is effective in inferring the input constraints.
Overall, \tool{} generates more crashing inputs than existing approaches.

\textbf{Contributions.} We present the following contributions:

\begin{itemize}
  \item To fuzz deep learning libraries, we propose that inputs can be categorized based on properties encoding domain knowledge of the libraries.
  This enables input constraint inference for supporting the generation of valid inputs and reducing redundancy in fuzzing the libraries.
  \item We design and implement our approach in \tool{}, 
  which employs active learning for input constraint inference. 
  During fuzzing, inputs are selected to answer queries related to a function's input constraints.
  \item Our evaluation shows that \tool{} improves over the existing approaches in finding crashing inputs. \tool{} generates more valid and diverse inputs. 
  Of the 43 new crashes found by \tool{}, 28 have been confirmed by the developers. The others are pending confirmation, or were already known but not yet fixed. 13 CVEs have been assigned so far. 
\end{itemize}

\section{Background}
\label{sec:background}

Deep learning libraries, such as TensorFlow and PyTorch, 
are employed by deep learning systems.
Library vulnerabilities widen the attack surface  
of the software systems that depend on them~\cite{xintan2022deeplearningsupplychain}. 
These vulnerabilities may, 
for example, allow denial-of-service attacks on software systems using them~\cite{tfsec}.

\textbf{Architecture.} The core functionality of the deep learning libraries is implemented in their kernels, which are written in a low-level language such as C/C++.
Applications using the libraries access their functionality through their Python API. The libraries perform validation on their inputs before accessing the core library code.

\textbf{Input domain of deep learning libraries.} Among others, the input domain of deep learning libraries includes tensors and matrices.
These inputs are complex; a tensor may be sparse or dense (corresponding to the format that the tensors are encoded), may be ragged (tensors with variable length), has a shape (dimensions of the matrix/tensor) and rank (number of linearly independent columns).
Functions in the libraries' APIs may impose constraints on its inputs, for example, requiring tensors of a specific type (e.g. float)
and size (e.g. a 3x3 matrix).
Xie et al.~\cite{xie2022docter} investigated TensorFlow's input constraints and categorized them by their \textbf{structure} (e.g., a list), \textbf{type} (e.g., tensor containing `float' values), \textbf{shape} (e.g., a 2-d tensor), 
and valid \textbf{values} (e.g., non-negative integers)

\textbf{Bugs in deep learning libraries.} Previous studies~\cite{jia2021symptoms,islam2019comprehensive,jia2020empirical} have empirically analyzed bugs in deep learning programs.
Jia et al.~\cite{jia2020empirical} reported that common root causes of bugs within TensorFlow include 
\textbf{type confusion} (incorrect assumptions about a variable type), \textbf{dimension mismatches} (inadequate checks of a variable's shape), and unhandled \textbf{corner cases} (usually related to incorrect handling of a specific variable's value, e.g. unhandled division by zero errors).
The overlap between the   root causes of bugs and input constraints suggests that distinguishing inputs by  these properties would help in both finding bugs and inferring input constraints.

\textbf{Testing deep learning libraries.}
Several recent works~\cite{pham2019cradle,wang2020deep,guo2020audee,wang2022eagle} mutate deep learning models 
for testing deep learning libraries.
Subsequently, the experiments of \texttt{FreeFuzz}~\cite{wei2022free} showed
that API-level testing of deep learning libraries is more effective.
\texttt{FreeFuzz} is seeded with inputs from publicly available code,
models, and library test cases.

To effectively test the deep learning libraries, the inputs selected by the fuzzer should be semantically valid, i.e. they should satisfy the input validation checks of the API function.
Otherwise, the core functionality of the libraries would not be tested. 
To address this, 
\texttt{DocTer}~\cite{xie2022docter} was proposed to exploit the libraries' consistently structured documentation to extract input constraints.
Still, as the API documentation is incomplete, manual annotation is required for part of the API and \texttt{DocTer} achieves a valid input generation rate of only 33\%.

\texttt{DeepRel}~\cite{deng2022fuzzing} and \texttt{FreeFuzz}~\cite{wei2022free}  
use seed inputs collected from publicly available resources (e.g., publicly available deep learning models, documentation, developer test suites).
\texttt{FreeFuzz} invokes functions in the API for which a valid invocation was observed from the resources.
Building on \texttt{FreeFuzz}, \texttt{DeepRel} generates valid inputs for some functions without seed inputs
by using the  {\em similarity} of pairs of functions to transfer inputs from test cases of similar API functions 
to other functions without seed inputs. 
The similarity of functions is determined based on the function signature and documentation, 
which may limit it to well-documented functions.
The existing approaches do not have a method of systematically selecting inputs to reduce redundancy.

\section{Preliminaries}



\begin{table}[t]
  \caption{A glossary of terms used in the Active Learning literature and this paper.}
    \label{tab:glossary}
    \begin{tabular}{ |l |}
      \hline
      \textbf{Active Learning}: An algorithm that learns by interactively\\
      ~~querying an oracle.\\
      \textbf{Consistency}: The extent to which executions under the  \\ 
      ~~inferred hypothesis matches the actual program \\
      \textbf{Input constraints}: The validation checks performed by   \\
      ~~the library on its inputs \\
      \textbf{Input properties}: Predicates which describe inputs \\
      \textbf{Input categories}: Conjunction of input properties. \\
      \textbf{Hypothesis}: A model of the input constraints as inferred \\ 
      ~~ by \tool{}. A disjunction of properties associated \\
      ~~  with a set of input categories.   \\
      \hline
  \end{tabular}
\end{table}

\begin{table*}[!t]
  \caption{Examples of input property templates. X refers to the input. C and C1 refer to constant values, which can be replaced with concrete values to instantiate a property. \tool{} uses a total of 92 property templates, which can be viewed on the artifact website~\cite{replication}.}
    \label{tab:properties}
    \centering
    \begin{tabular}{ |l |l | l|}
      \hline
      \textbf{Property group} & \textbf{Example} & \textbf{Description} \\ 
      \hline
      Type/Structure  &  isinstance(X, type) & the type of the input  (e.g. a list)  \\
      
      &  X.dtype = type & the type (e.g. int) of elements in a tensor/matrix   \\
      \hline
       &  X $<$ C & ranges of values \\
      Value & all(X $>$ C) & ranges of values of elements in a tensor/data structure \\
                        
       &  X[C] = C1 & value of a specific element \\
      \hline
       & len(X) $<$ C & length/size of a data structure \\
      Shape & X.shape.rank $>$ C & rank of a matrix \\
       & X.shape[C]  == C2 & size of a specific dimension \\
      \hline
  \end{tabular}
\end{table*}

\subsection{Active Learning}
We apply active learning for  input constraint inference. 
To infer and refine a model of a function's input constraints, our approach generates inputs that provide  new information when they are used to invoke a function.
Table \ref{tab:glossary} presents a glossary of terms used in the active learning phase of \tool{}.

In active learning~\cite{angluin1987learning,angluin1988queries}, a learner sends queries to an oracle who responds with some feedback (e.g., the ground truth label of a given data instance).
When active learning is employed for inferring a model of a program, a \textit{hypothesis} is a possible model.
A hypothesis is \textit{consistent} if the behavior expected from the model matches the actual behavior of the  program.

In this paper, active learning is done while fuzzing the deep learning libraries.
Our study combines active learning with fuzzing to learn the input constraints of an API.
Fuzzing is used to learn a model of the input constraints, which are, in turn, used to improve fuzzing by enabling the generation of semantically valid inputs.

\textbf{Input constraint inference.} Our approach, \tool{}, aims to infer accurate models of the \textit{input constraints} of the functions in the deep learning libraries' API.
Input constraints refer to the conditions on the inputs that are expected to be fulfilled for the function to be successfully invoked.
Code in the library typically performs validation checks on the inputs, ensuring that the constraints are satisfied before executing the core functionality of the library.
\tool{} refines a \textit{hypothesis} of the input constraints of the API functions  as \textit{queries} are made to the test executor.
The test executor answers the queries by checking if an input with properties corresponding to the query satisfies the input constraint (i.e., if it is \textit{valid}, \textit{invalid}, \textit{crash}) determined by observing if the function was invoked without error (\textit{valid}), rejects the input through an exception (\textit{invalid}), or crashes the program.
A crashing input is one that causes the library to terminate in an unclean manner (e.g. segmentation faults), which leads to a denial-of-service.

\subsection{Input properties}

\tool{} characterizes inputs to the deep learning libraries
by \textit{input properties}.
The properties are used by \tool{} to distinguish inputs from one another.
Some examples of the properties are given in Table \ref{tab:properties}.
The properties were designed based on previous empirical studies of deep learning libraries, 
which found that the common root causes of bugs are \textbf{type confusion}, \textbf{dimension mismatches}, and unhandled \textbf{corner cases}. 
The root causes motivate properties related to an input's \textit{type}, \textit{shape}, and \textit{value}, respectively.

\begin{figure}[t]
  \small
\begin{lstlisting}[language=Python]
# Input 1
# property: type(X) == RaggedTensor
input1 = tf.ragged.constant([[1,1,1,1], [2]])

# Input 2
# property: X.dtype = [('qint32', '<i4')] 
input2 = tf.constant(np.zeros((1,1,1,1), 
  [('qint32', '<i4')]))

\end{lstlisting}
\caption{Example of inputs constructed using \texttt{tf.constant} differing in their  properties}
\label{fig:example1}
\end{figure}

Two example inputs are shown in Figure \ref{fig:example1}. 
These inputs, which are constructed using \texttt{tf.constant}, are both \texttt{Tensors}.
However, they both satisfy at least one property that is not satisfied by the other.
As they do not share the same properties, they are more likely to trigger different behaviors when passed to the same function.

\subsection{Input categories}
\label{sec:input_cats}

As each input can satisfy multiple input properties,
\tool{} characterizes each input with  the properties that it satisfies.
As a pre-processing step of fuzzing, \tool{} groups inputs that satisfy the same properties.
An \textit{input category} is a conjunction of input properties and is associated with the inputs that satisfy the conjunction of properties.

As all inputs in a category satisfy the same conjunction of properties, 
they all satisfy the input constraints corresponding to the properties.
For example, an input of a category with the property \texttt{X.shape.rank == 4} will pass the validation checks of a function requiring an input of rank 4.

\begin{definition}
  Two inputs, $x$ and $y$, \textbf{belong to the same input category}, $C$, if every  property that  $x$ satisfies  
  matches a  property that $y$ satisfies, and vice versa.
  \end{definition}

A category contains the inputs that satisfy the same properties
We assume that the true input constraints of the function corresponding to a set of input categories, i.e., a collection of properties describing valid inputs.  
For example, all inputs in the input category associated with \{\texttt{X is not None}, \texttt{X.shape = (2,2)} \} are tensors of the same shape and will satisfy input constraints requiring tensors of this shape.
The execution of multiple test cases 
selecting inputs from different categories provides information about the function's true input constraints.
As such, the input categories allow the systematic selection of inputs during fuzzing.

\begin{definition}
  An input category, $C_1$, is \textbf{weaker} than an input category, $C_2$, if the set of inputs associated with $C_1$ is a superset of the set of inputs associated with $C_2$.
  \end{definition}

\tool{} orders the categories by their strengths. 
One category, $C_1$, is stronger than  another if it has input properties that are associated with inputs 
that are a subset of the inputs associated with the other category, $C_2$.
Intuitively, if  the input constraints of a function match the properties of an input category, $c_1$, 
then we assume that inputs from a stronger category, $c_2$, would be  valid.
Inputs from the stronger category will observe the same input properties of the weaker category, and will satisfy the corresponding input constraints.

For example, given a first category associated with the set \{\texttt{X is not None}, \texttt{X.shape = (2,2)} \},
and a second category associated with \{\texttt{X is not None}, \texttt{X.shape = (2,2)},  \texttt{$\forall$ x $\in$ X (x > 0)} \}, the first category is weaker than the second category (as the first category has fewer properties to satisfy). 
If the first category is already a match for the actual input constraints, then we expect that the inputs from the second category would be valid.

\tool{} maintains a mapping of the input categories to the inputs satisfying the associated properties.
This enables it to quickly sample the inputs during fuzzing.
The input categories also allow for the sampling of fewer redundant inputs. 
An input sampled from an input category will satisfy all properties associated with the input category.
To obtain some evidence that inputs from a category satisfy the actual input constraints of a function, 
\tool{} observes 
the outcome of invoking the function with inputs sampled from the category.

\begin{definition}
 A \textbf{hypothesis} is a disjunction of properties associated with a set of input categories.
  \end{definition}

\tool{}  models the input constraints of a function as a set of input categories.
The active learner infers and refines its \textit{hypothesis} of the true input constraints expressed as a disjunction of the properties associated with the input categories. 
A disjunction of input categories is used because the functions in TensorFlow and PyTorch allow for a union of types, a common feature of dynamically typed languages, e.g., Python.
For example, a hypothesis can be constructed by the selection of two input categories, one associated with the set 
\{\texttt{X is not None}, \texttt{X.type = list}, \texttt{len(X) = 4} \}, and another with the set \{\texttt{X is not None}, \texttt{type(X) == Tensor)}, \texttt{X.dtype == tf.int64} \}.
The hypothesis captures a different set of properties depending on the input's type.
Once \tool{} considers a hypothesis of the input constraints to be adequate, \tool{} then generates test cases using inputs  expected to be valid according to the inferred input constraints.

\subsection{Motivating Example}

\begin{figure}[t]
  \small
  \begin{lstlisting}[language=Python]
import tensorflow as tf
  
# generate inputs
input1 = tf.constant([1, 2, 3])
shape = [4]

# invoke the target API function with 
# the generated inputs
tf.placeholder_with_default(input1, shape)
  \end{lstlisting}
  \caption{Example of a test input generated for \texttt{placeholder\_with\_default}. Inputs for each argument (e.g. \texttt{shape}) are generated. 
  A valid input of \texttt{shape} can be a \texttt{tf.TensorShape} \textbf{or} a \texttt{list} of \texttt{int}
  }
  \label{fig:examplecrash}
  \end{figure}
  
Figure \ref{fig:examplecrash} shows an example of a test case generated for the API function, \texttt{tf.placeholder\_with\_default}.
To generate semantically valid inputs that satisfy the function's input constraints, the inputs require the right type and satisfy other constraints on their shape and values. 
If \texttt{shape} is a list, there are other constraints such as the type or range of values of its elements.
If provided an input that does not meet these constraints, the library signals that the input is \textit{invalid} by throwing an appropriate exception.

\begin{figure*}[!t]
  \includegraphics[width=0.9\textwidth]{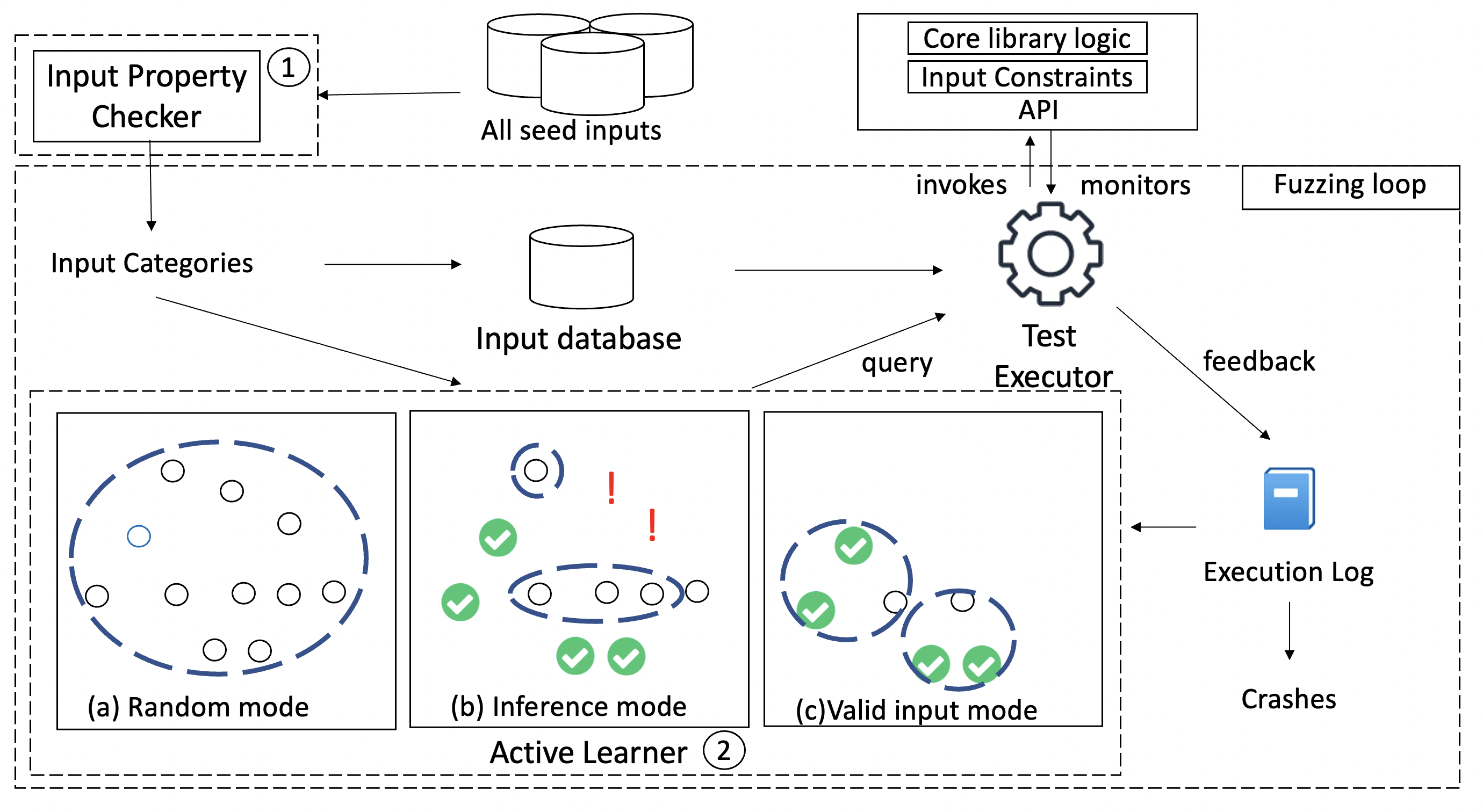}
\centering
\caption{Overview of \tool{}. Step \circleone{}: Seed inputs (e.g.  collected from the library's test suite) are grouped into input categories. Step \circletwo{}: As the library is fuzzed, the active learner has 3 modes that determine what queries are posed to the test executor. 
It selects input categories from the input space. Initially, in its (a) \textit{random generation} mode, it randomly selects input categories (denoted as circles). Next, in its (b) \textit{inference} mode, it selects input categories to refine its hypothesis of the input constraints, which are formed based on the execution log which indicates the outcomes (check marks denote \textit{valid} inputs and exclamation marks (!) denote \textit{invalid} inputs) of prior queries. Finally, once a hypothesized input constraint is accepted, in its (c)  \textit{valid input generation} mode, it selects only inputs satisfying the hypothesis. Each blue, dashed ellipse shows the narrowing space of categories considered in each mode. }
  \label{fig:overview}
\end{figure*}

To generate valid inputs, \tool{} has to discover the input constraint by invoking the function multiple times with different values of \texttt{shape} and observing the result of each invocation.
A successful invocation indicates that the input satisfies the input constraints, and an unsuccessful invocation indicates otherwise.
\tool{} forms a hypothesis regarding the constraints of \texttt{shape}.
As previously described, \tool{} expresses a hypothesis as a disjunction of properties so that it can capture input constraints that are a union of multiple constraints.
The true constraints of the \texttt{shape} parameter permits inputs typed \texttt{list} or \texttt{tf.TensorShape}. 
\tool{} has to express one set of properties if provided a \texttt{list} and another set of constraints if provided a \texttt{tf.TensorShape}.
 
Once the input constraints are successfully inferred, \tool{} generates inputs that are \textit{valid},
i.e., invoking the function without error, by sampling inputs from the input categories in the hypothesis.
This allows \tool{} to generate inputs that pass the input validation checks and test the core functionality of the library.
Testing the libraries with a diverse range of inputs is key to finding crashes.
In Figure \ref{fig:examplecrash}, if  \texttt{shape} is a quantized tensor, then the library's kernel code does not correctly access its memory contents and will trigger a segmentation fault.
In other words, a quantized \texttt{shape} is a crashing input.
Finding a crashing input poses a challenge as the space of inputs is large and there are only a few crashing inputs. 
Many inputs are redundant as they share the same properties. 
For example, all inputs with the same wrong shape will fail the same validation check on the input shape and not reach the core functionality of the library.
To this end, \tool{} does not get stuck with inputs that fail the same validation checks as using them does not provide \tool{} with new information. 
Instead, \tool{} skips past the inputs in the same category to inputs from other categories, invoking the function with more informative inputs.

\section{\tool{}}
\label{sec:approach}

\subsection{Overview}

Figure \ref{fig:overview} shows the overview of \tool{}.
In the first step (\circled{1} in Figure \ref{fig:overview}), 
\tool{} collects inputs from the execution of the library's test suite and associates them with properties that they satisfy (Section \ref{sec:input_collection}).
Then, each input is grouped into input categories with other inputs satisfying the same properties.
These inputs form the input space considered by \tool{}.
In the second step (\circled{2} in Figure \ref{fig:overview}), \tool{} fuzzes the deep learning libraries.
This involves the generation of test cases by selecting inputs to use as arguments in invoking the API functions.
The selection of inputs involves an active learning algorithm that infers the input constraints of a target API function.
The active learner constructs queries to check if an input category is a member of the input domain, i.e., its inputs are valid.
The test executor has the role of the oracle; to respond to the query, it invokes the library with appropriate inputs sampled from the queried category, checking if they satisfy the actual input constraints (i.e., the function invocation does not lead to an exception or crash).

The test executor constructs test cases by sampling inputs associated with the target input categories.
As it constructs and executes a test program, 
the invocation of the library is monitored for  crashes (errors in the C++ code of the libraries that may be exploited by an attacker, e.g., segmentation faults) 
and exceptions thrown by the library are caught.
The observation (i.e., query and the outcome of the test execution, \textit{valid}, \textit{invalid}, or \textit{crash}) is written to the execution log.
Considering these observations, the active learner refines its hypothesis and constructs more queries.

During fuzzing, 
\tool{} employs active learning to learn the input constraints of the API function. 
The fuzzing loop involves an active learner and a test executor.
The active learner maintains a hypothesis of the input constraints of a given API function.
To check the hypothesis, it passes queries to the test executor.
Each query is one input category.
On receiving the query, the test executor samples an input that satisfies the input category and constructs a Python program that invokes a function from the library's API.
Each constructed Python program consists of code that generates the inputs (e.g., the variable, \texttt{shape}, in Figure \ref{fig:examplecrash}) using 
program fragments (e.g., invocation of \texttt{tf.constant}) collected from the developer test suite.
After the inputs are selected, they are passed as arguments to the API function under test (e.g., \texttt{tf.placeholder\_with\_default}).

When fuzzing each target function, there are three phases (described in detail in Section \ref{sec:fuzzing}).
Initially, as there is no history to support a hypothesis, \tool{}'s randomly selects input categories from the entire input space ((a) in Figure \ref{fig:overview}).
Afterward, the active learner begins to pose queries to the test executor for input constraint inference (described in Section \ref{sec:active_learning}).
These queries are selected based on the hypotheses ((b) in Figure \ref{fig:overview}).
Each query corresponds to one input category.
Finally, once the hypothesis is determined to be adequately consistent, then \tool{}  selects only inputs that satisfy the input constraints indicated by the hypothesis ((c) in Figure \ref{fig:overview}).

\tool{} maintains a list of crashing test cases.
When \tool{} is terminated, the crashes are the output of \tool{} and can be inspected.

\subsection{Step 1: Input property checking and input category construction}
\label{sec:input_collection}

\tool{} requires seed inputs before it begins categorizing them.
In our experiments, we use the developer test suite, which is readily available from the deep learning libraries' repositories, and execute them to obtain seed inputs.
Before the execution of the test cases, \tool{} instruments the API functions.
As the test cases are executed, 
the inputs passed as arguments to the functions of the APIs are traced.
Whenever a library test case invokes the API function (either directly or transitively), 
the API function sequences (e.g. $\langle$ \texttt{tf.constant}, \texttt{tf.ragged.constant} $ \rangle $) for generating the argument inputs are recorded.
This enables \tool{} to reconstruct the inputs used in the developer test suite.

\tool{} enumerates the possible properties for each obtained  input,
checking if the input satisfies the input properties.
After associating the satisfied properties with every input, \tool{} groups them into input categories.
The input categories are fixed at this time, to be later used during fuzzing.
A mapping from categories to their inputs is maintained by \tool{} for efficient sampling of the inputs.

\textbf{Reducing input redundancy.} 
\tool{} leverages the input categories to reduce redundancy.
As inputs from the same categories share the same properties, they  will satisfy the same  constraints corresponding to these properties.
By selecting inputs from different categories, \tool{} aims to not construct multiple test cases  with similar inputs that will fail the same validation checks,
as it does not gain information necessary for refining its hypotheses.
By avoiding the use of inputs that are similar to previously selected inputs,
each test case is more likely to provide new information about the true input constraints.
Hence, using inputs from different categories leads to the use of fewer redundant inputs.

\subsection{Step 2: Active Learning-driven fuzzing}
\label{sec:fuzzing}

In the second step, \tool{} begins fuzzing the deep learning libraries. This is done through three phases.

\textbf{(a) Random inputs generation.}
\tool{} begins generating test cases for each target function by selecting inputs from random  categories.
This phase ends once \tool{} successfully generates a test case with valid inputs (i.e., the function executes without error using the inputs).

\textbf{(b) Input constraint inference.}
Once a valid input has been identified, \tool{} is able to form hypotheses of the input constraints (later described in Section \ref{sec:active_learning}).
\tool{} tests the hypothesis that is most consistent with the observations by selecting queries based on the hypothesis.
It selects input categories from which inputs should be valid according to the hypothesis, as well as categories from which invalid inputs should be produced.
Through interacting with the test executor, the active learner refines the hypothesis.

The active learner forms hypotheses of the correct input constraint, assessing them by quantitative measures of consistency.
These measures are computed using the number of observed valid and invalid inputs that correctly and incorrectly satisfy the hypothesized input constraints.

\textbf{(c) Valid input generation.}
Once \tool{} considers a hypothesis adequately consistent,
\tool{} begins to construct test cases with inputs that are valid according to the hypothesized input constraints.
This is done by sampling inputs from the input categories that are part of or are stronger than the hypothesis.

\begin{algorithm}[t]
  \SetAlgoLined

  \nonl \Function fuzz($\mathit{input\_categories}$):

    $history$ = []

    \While{not done}{
       $selected\_category \leftarrow $ active\_learner($input\_categories$,$history$)

       \eIf{selected\_category == null}{$random\_category$ = sample($input\_categories$)

          $input \leftarrow $ sample($random\_category$)
       }{$input \leftarrow  $ sample($selected\_category$)
       }

       $tc  \leftarrow $ construct\_test\_case($input$)

       $outcome \leftarrow $ execute($tc$)

       update($history$, $selected\_category$, $outcome$)
  }
  
   \caption{The fuzzing loop of \tool{} which involves the active learner posing queries to the test executor.}
   \label{fig:fuzz}
\end{algorithm}

The procedure for selecting one argument given an API function is given in Algorithm \ref{fig:fuzz}.
Initially, \tool{} begins the fuzzing campaign with purely random inputs as the active learner is not able to construct queries without  previously observed executions (lines 4--6). 
After one valid input is observed, the active learner begins to pose queries to the test executor, which constructs test cases based on the queries (line 3).
When \tool{} has entered its valid input generation mode, the active learning only poses queries to guide the selection of inputs that are expected to be valid according to the hypothesis.
Given the input category in a query, the fuzzer selects a random input that is associated with the input category (line 8).
With the selected input, a test case is constructed and then executed to invoke the library (lines 10--11).
The outcome of the test execution is written to the execution log (line 12), $history$, which is used in the next iteration by the active learner to pose a new query.

\subsection{Input constraint inference}
\label{sec:active_learning}

The key novelty of \tool{} is that it learns the input constraints while fuzzing the API function (in (b) of step \circled{2} in Figure \ref{fig:overview}). 
Through the interaction of the active learner with a test executor, the active learner records the test outcomes in the execution log.
These observations are used to form and refine hypotheses of the input constraints, and for the active learner to pose queries to the test executor.

\textbf{Selecting queries based on a hypothesis.} We refer again to Table \ref{tab:glossary}, the glossary of terms used in the active learning phase of \tool{}.
The active learner in \tool{} poses \textit{queries} to the test executor to check if its \textit{hypothesis} of the actual input constraints indeed match  the true input constraints.
If the hypothesis is a match, then inputs satisfying the hypothesis should be accepted by the library while inputs that do not satisfy the hypothesis should be rejected by the library.
Hence, we expect that inputs from the input categories of the hypothesis should lead to \textit{valid} outcomes.
Conversely, inputs that are missing at least one property in a category of the hypothesis should be rejected.
We expect that these queries should result in \textit{invalid} outcomes.
If these queries lead to valid outcomes, then it implies that the hypothesis is stronger (see Definition 2 in Section \ref{sec:input_cats}) than the true input constraints.

As such, for one hypothesis, the active learner constructs several queries by selecting input categories with respect to the input categories that compose the hypothesis.
One set of queries checks that inputs satisfying the hypothesis also indeed satisfy the actual input constraints (i.e., the test constructed will be executed without error).
Another set of queries checks if the inputs that do not satisfy the hypothesis also do not satisfy the input constraints (i.e., the test constructed results in an error when executed).

At the beginning of the fuzzing campaign, \tool{} selects random inputs.
As  the fuzzing proceeds, the active learner begins to pose queries to the test executor.
The active learner considers  the execution log to select input categories to form a hypothesis, and selects input categories as queries. 
It optimizes for the confirmation of possible hypotheses of the input constraints of the API function.

As the test executor component evaluates a test case, the query (i.e., choice of input category) and test outcome are written in the execution log. 
If a test case results in an exception thrown by the deep learning library, then the input is \textit{invalid}.
If the library invocation succeeds without any exceptions, then the input is \textit{valid}.
If the test case crashes the deep learning library, then the input is a \textit{crash}ing input.

\textbf{Measuring consistency.} At any given time, there may be multiple hypotheses that can be considered by the active learner.
The active learner selects the hypothesis that is the most consistent with the observations in the execution log.
To do so, it quantitatively measures the number of valid and invalid inputs that are consistent with the hypothesis.
Given a perfectly consistent hypothesis, all valid inputs will be included in an input category in the hypothesis.
Conversely, all invalid inputs should not be a member of the hypothesis.

As it may not be possible to infer a perfectly consistent hypothesis,
we compute quantitative measures of a hypothesis' \textit{consistency}. 
Each hypothesis proposed by the active learner is assessed on its consistency with regard to the observations in the execution log;
valid inputs should satisfy the input constraints in the hypothesis and invalid inputs should not.
Within \tool{}, we do not expect that the hypothesis will perfectly match the input constraints. 
As such, \tool{} assesses each hypothesis and selects one that is the most consistent with the observed executions.
A good hypothesis includes input constraints that cover a large part, if not all, of the valid observations.
It should also not incorrectly cover invalid inputs.
\tool{} uses precision and recall  to assess the quality of a hypothesis.
Out of \texttt{all} inputs selected, given that 
\texttt{covered(valid, hypothesis)} represents the number of valid inputs that fall within the hypothesis,
and \texttt{covered(all, hypothesis)} represents the number of inputs, both valid and invalid, that fall within the hypothesis.
The precision, P, and recall, R, are computed as follows:

\vspace{-0.2cm}
\begin{align*}
P =  \frac{covered(valid, hypothesis)}{covered(all,hypothesis)} 
\end{align*}
\vspace{-0.2cm}
\begin{align*}
R  =  \frac{covered(valid, hypothesis)}{ | valid | } 
\end{align*}

Precision measures the proportion of valid inputs that fall within the hypothesized input constraints out of all the observed 
inputs.
Recall measures the proportion of valid inputs that fall within the hypothesized input constraints out of all the observed valid inputs.
Together, the two metrics measure the adequacy of the hypothesized input constraint.
A hypothesis is adequately consistent if the precision and recall exceed a threshold set at the start of the fuzzing campaign.

\section{Implementation}
In the previous section, we have discussed the key ideas behind \tool{}. Here, we discuss the implementation details.

\textbf{Building the input database.}
\tool{} is implemented as a Python program that takes the API 
and the developer test suite as its input.
The list of functions in the API is obtained.
We obtain the input values used in the library test suite as the seed inputs for \tool{}, the Python library code is instrumented to track the invocation of every function call to record their argument inputs.
The functions to construct the inputs, the returned values of their invocations, and the input properties satisfied by the inputs are stored in the database.
Inputs are generated by fetching and invoking the functions.

\textbf{Crash Oracle.}
As our research objective in this study is to assess the ability of \tool{} to explore the input space, 
we only monitor the deep learning libraries for crashing inputs.
\tool{} is implemented with a crash oracle.
The test executor constructs and executes a test program on a different process. 
Then, the test process is monitored for crashes.
Inputs that crash the library are written to the execution log.
These crashes are later investigated manually to identify unique crashes before we report them to TensorFlow and PyTorch.

The crash oracle detects weaknesses considered as security vulnerabilities~\cite{tfsec} (e.g. segmentation faults) that cause the running process to terminate in an unclean way.
Other methods of detecting vulnerabilities may be implemented in \tool{} in the future,
but in our experiments, 
we focused on uncovering crashes in the libraries that may be exploited for denial-of-service attacks.

\textbf{Active Learning.}
The active learner takes the execution log as input and produces a series of queries to be posed to the test executor.
The  queries are constructed  based on the subset of input categories in the hypothesis.
The construction of a hypothesis and the selection of queries are obtained through the execution of a logic program.
Using a logic program allows us to declaratively express the desired characteristics of a hypothesis and optimize  the selection of input categories against a criteria.
The active learner selects an appropriate hypothesis while maximizing  
the number of valid inputs that match the hypothesis, 
minimizing the number of invalid inputs that are incorrectly matched by the hypothesized input constraint,
and favouring simpler hypothesis by minimizing the number of input categories used in the hypothesis.
In this way, \tool{} assesses each hypothesis on its consistency with the observed test outcomes.

\tool{} accepts a hypothesized input constraint considering if its precision and recall exceed a threshold.
In our experiments, we set a low threshold for both precision and recall at 0.25.
This enables the input constraints to be inferred for a large proportion of the API. 
As our goal is to fuzz the API thoroughly, we find allowing the fuzzer to focus on a broad region of inputs that include the valid domain of inputs of the functions is more beneficial than precisely identifying the valid domain of inputs.
We empirically find that the low thresholds do not adversely impact the 
proportion of valid inputs selected by \tool{} when using the hypothesized input constraints to generate valid inputs.
This is because the logic program already optimizes the selection of hypotheses for a high level of consistency.

\textbf{Interleaving of target functions.}
The active learner \tool{} employs \texttt{clingo}~\cite{gebser2008user} to execute the logic programs used by \tool{} to select the next set of inputs.
Logic programs take a significant amount of time to be executed to produce their output.
To allow time for the logic program to be executed, 
\tool{} interleaves the construction of test cases for different API functions,
coming back to the same function only after completing a test case for each of the other test cases.
This provides ample time for the logic program to be run before the same API function is tested again.

\section{Evaluation}
\label{sec:evaluation}

\subsection{Research Questions}

Our experiments aim to provide answers to the following research questions.
We investigate \tool{} capability in finding crashing inputs (RQ1).
Next, we analyze the input generation ability of \tool{} (RQ2 -- RQ4).

\vspace{0.2cm}\noindent{\bf RQ1. Does \tool{}  produce crashing inputs?} 

This question concerns the ability of \tool{} in triggering crashes, which is our primary objective.
We count the number of new crashes that have not been previously reported, which we then reported to the library developers for validation.
We also compare the ability of each approach in triggering the set of crashes found by at least one approach.

\vspace{0.2cm}\noindent{\bf RQ2. Does \tool{} sample diverse inputs? } 

Active learning should enable \tool{} to reduce redundancy during fuzzing by selecting a wide range of input categories.
We investigate if \tool{} was able to do so.
We compare \tool{} against the baselines and 
compare the inputs generated to fuzz the functions known to crash.

\vspace{0.2cm}\noindent{\bf RQ3. Does \tool{} sample valid inputs?} 

\tool{}  is expected to generate a larger proportion of valid inputs.
We investigate if inputs selected to satisfy the inferred input constraints are indeed valid inputs.

\vspace{0.2cm}\noindent{\bf RQ4. Which components of \tool{} contribute to its ability to find crashing inputs?} 

\tool{} aims to have a less redundant selection of inputs and generate a higher proportion of valid inputs.
We perform an ablation study to determine how the  components of \tool{} contribute to it.

\subsection{Experimental Setup}

\textbf{Baselines.}
We compare \tool{} against the state-of-the-art deep learning library fuzzers targeting the libraries' API, \texttt{DeepRel}~\cite{deng2022fuzzing} and \texttt{DocTer}~\cite{xie2022docter}.
We run the tools from their replication packages and analyze the list of bugs reported.

\texttt{DeepRel} builds on top of \texttt{FreeFuzz}~\cite{wei2022free}, using the same strategy of generating inputs for each API function. 
\texttt{DeepRel} and \texttt{FreeFuzz} collect inputs for use from open-source code on GitHub, publicly available models, and the library test suite.
Compared to \texttt{FreeFuzz} and \texttt{DeepRel}, \tool{} uses only inputs from the libraries' test suites while 
\texttt{DeepRel} and \texttt{FreeFuzz} use seed inputs collected from open source resources.
As \texttt{DeepRel} and \texttt{FreeFuzz} uses the same input generation strategy and differ only in the number of API functions they cover,
we only compare \tool{} against \texttt{DeepRel}.

\texttt{DocTer} extracts input constraints from the library documentation.
Then, it generates inputs to invoke the libraries    considering the extracted input constraints.

\textbf{Environment.}
We run experiments on TensorFlow 2.7.0 and PyTorch 1.10, the same version of the libraries used by the most recent study~\cite{deng2022fuzzing}.
We collect a list of all API functions of TensorFlow and PyTorch.
It is used in our initial experiments, where we attempt to run the approaches on every function.
Subsequently, we focus our analysis on the ability of the fuzzers to trigger the crashes found by the approaches. 
Using \texttt{DocTer}, \texttt{DeepRel}, and \tool{},
there are crashing inputs to 231 functions in TensorFlow and 
95 functions in PyTorch.

We configured and ran the fuzzers for up to 48 hours.
In the prior experiments of the baseline fuzzers~\cite{xie2022docter,wei2022free,deng2022fuzzing}, the tools were allowed 
up to 1,000~\cite{wei2022free, deng2022fuzzing} or 2,000~\cite{xie2022docter} executions for each function.
To generate 1,000 test cases, we executed \texttt{DeepRel} and it took 172 hours and 43 hours to complete generating test cases for TensorFlow and PyTorch.
\texttt{DocTer} took 16 hours for TensorFlow and 25 hours on PyTorch.
Therefore, to use the same budget for a fair comparison, we tweaked the number of test cases generated by the baseline fuzzers to fit in 48 hours and reran the fuzzers.

Our experiments on executed on a machine with an Intel(R) Xeon(R) CPU E5-2640 v4 @ 2.40GHz, 205G, Tesla P100. 
While our fuzzer does not directly use the GPU, some functions in the library may use the GPU.

\subsubsection{Evaluation Metrics.}

We use the following metrics to assess \tool{}:

\begin{itemize}
    \item \textbf{\# of detected crashes.} The primary goal of \tool{} is to generate inputs to crash TensorFlow and PyTorch.
    \item \textbf{Input property coverage.} 
We report the number of unique input properties that have been satisfied by at least one input during fuzzing. 
To reduce input redundancy, a high input diversity is desirable.
To measure the diversity of inputs, we count 
the total number of unique input properties observed to be satisfied at least once.
    \item \textbf{API coverage.}
We report the number of functions that valid inputs were successfully generated for.
This metric was previously used in the evaluation of \texttt{DeepRel}~\cite{deng2022fuzzing}.
Related to this metric, we also report the proportion of generated valid inputs.
\end{itemize}

\subsection{Experimental Results}

\subsubsection{RQ1. Vulnerabilities detected}

\begin{table}[t]
  \caption{The number of unique new vulnerabilities of TensorFlow reported in this study and prior studies.}
    \label{tab:new_vulns}
    \begin{tabularx}{\linewidth}{ |X | r| r| r|}
      \hline
      \textbf{Approach} & \textbf{\# new vulnerabilities}  \\
      \hline
      \texttt{DocTer} &  1 \\
      \texttt{FreeFuzz} & 7 \\
      \texttt{DeepRel} & 1 \\
      \tool{} & 23 \\
\hline
  \end{tabularx}
\end{table}

\textbf{Existing crashes.} 
We perform a thorough analysis of the capability of the approaches in detecting existing crashes.
In this analysis, we consider all crashes found by the approaches.
To perform this analysis, we consider that a vulnerability was not detected if its corresponding API function
is not covered by the tool or if the tool does not report the bug although test cases were generated for the function.

In total, \tool{} detects a total of 168 crashing functions, 108 in TensorFlow version 2.7.0 and 58  in PyTorch version 1.10.
From the 108 TensorFlow functions, we grouped related crashes and reported 43 vulnerabilities.
From the 58 PyTorch functions, we reported 10 vulnerabilities.
After corresponding with the developers of TensorFlow and PyTorch, they confirmed that 23 of the TensorFlow vulnerabilities and 5 of the PyTorch vulnerabilities were previously unknown.
The remaining crashes are confirmed as vulnerabilities too, but they were already known by the developers (although the fix was not released yet).

We received 13 CVEs from these reports, with another 10 already confirmed. 2 are pending confirmation.
Next, we analyze the extent to which \tool{}, \texttt{DocTer}, and \texttt{DeepRel} are able to detect the same vulnerabilities.

\texttt{DocTer} found  163 
crashing functions.
Of the 108 vulnerable TensorFlow and 58 vulnerable PyTorch functions found by \tool{}, 
\texttt{DocTer} was able to 
successfully generate crashing inputs to 6 of the 108 vulnerable functions in TensorFlow.
and 12 of the 58 vulnerable functions in PyTorch.
Overall, \texttt{DocTer} detects just 18 of the 166 vulnerable functions detected by \tool{}.

On the other hand, when executed on the versions of libraries before these crashes were fixed, \tool{} is able to detect 52 (84\%) out of the 62 crashing functions in TensorFlow detected 
by \texttt{DocTer}.
On PyTorch, \tool{} is able to detect 7 (23\%) out of the 31 crashing functions detected 
by \texttt{DocTer}.
Overall, \tool{} detects 59 (63\%) out of 93 crashing functions detected by \texttt{DocTer}.

\begin{figure}[!th]
  \includegraphics[width=0.5\textwidth]{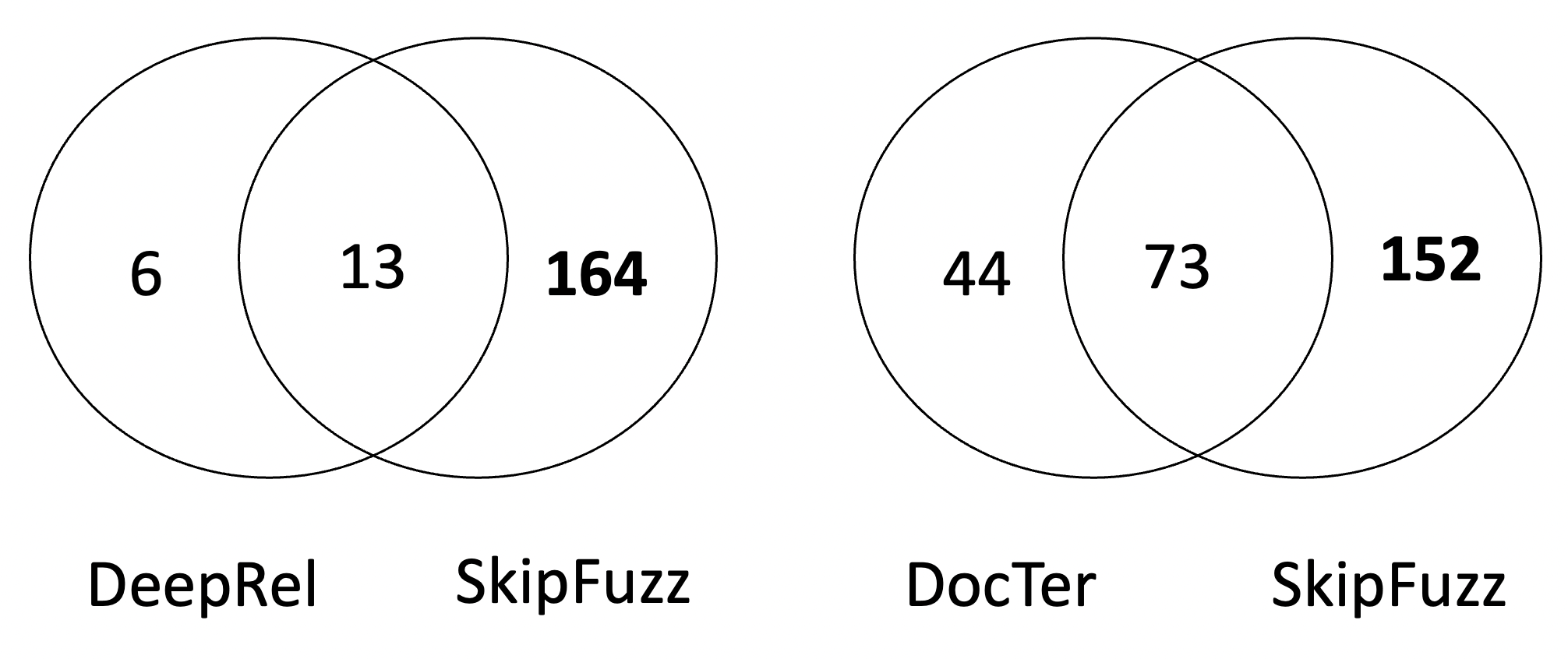}
\centering
\caption{Crashes in the deep learning libraries found by \texttt{DeepRel}, \texttt{DocTer}, and \tool{}.}
  \label{fig:venndiagram}
\end{figure}

Next, we compare \tool{} against \texttt{DeepRel} and \texttt{FreeFuzz}.
\texttt{DeepRel} and \texttt{FreeFuzz} was able to detect crashes for only 9 of the 108 TensorFlow functions and 7 of the 59 PyTorch functions.
The original experiments done to evaluate  \texttt{FreeFuzz}~\cite{wei2022free} and \texttt{DeepRel}~\cite{wei2022free} resulted in 
39 bug reports
on TensorFlow, of which 
10
involved crashes, and 
72 bug reports
on PyTorch, of which 
7
involved crashes.
When executed on versions of the libraries before the crashes were fixed, \tool{} is able to detect 8 (80\%) of the 10 crashes on TensorFlow and 3 (43\%) of the 7 crashes on PyTorch.
Figure \ref{fig:venndiagram} shows two Venn diagrams of the functions that each approach is able to generate crashing inputs to.

\textbf{New vulnerabilities in TensorFlow.}
Table \ref{tab:new_vulns} shows the number of new crashes found in the experiments of \texttt{DocTer}, \texttt{FreeFuzz}, and \texttt{DeepRel}.
On TensorFlow, 
we determine if a crash is new 
by going through the list of TensorFlow vulnerability reports
and comparing the referenced bug reports 
against the bug reports referenced by the replication packages of the prior approaches.
\texttt{DocTer}~\cite{xie2022docter} found 1 newly discovered vulnerability.
\texttt{FreeFuzz}~\cite{wei2022free} and \texttt{DeepRel}~\cite{deng2022fuzzing} found a total of 
8
crashes.
In our experiments, \tool{} detects 33 new vulnerabilities.
23 of them have been confirmed by TensorFlow developers to be new vulnerabilities, with
13 CVE IDs assigned. 
2 of them are pending confirmation.
We do not perform this analysis for the crashing inputs to PyTorch as its developers do not 
assign CVEs to potential security weaknesses.

There has been significant effort in detecting TensorFlow vulnerabilities.
Apart from the baseline approaches discussed,
it is a fuzz target in the OSS-Fuzz project~\cite{ossfuzz}.
OSS-Fuzz has found over 30,000 bugs in open source projects, including 6 security bugs in TensorFlow\footnote{Issues tagged  ``Bug-Security'' on \url{https://bugs.chromium.org/p/oss-fuzz/issues/list?sort=-opened&can=1&q=proj:TensorFlow}}.
Evidently, finding new vulnerabilities is not trivial.

\subsubsection{RQ2. Reducing input redundancy.}

To investigate the factors contributing to \tool{}'s performance,
we study the inputs used in fuzzing. 
We evaluate the reduction in redundancy by measuring input diversity.

We analyze the coverage of input properties by the inputs generated by the approaches. 
A higher coverage of input properties indicates a greater diversity of inputs. 
This suggests a low amount of redundancy in input generation.
Conversely, a low coverage may indicate that similar inputs was generated over and over again, which implies a high level of redundancy as the inputs may be failing the same validation checks, or triggering the same library behavior.
We investigate the number of properties that were satisfied by an input passed to TensorFlow's and PyTorch's API 
as a proportion of all input properties observed in the experiments.
In this analysis, we focus on the test cases generated to target the crashing functions to investigate the reason for \tool{}'s stronger ability to generate crashing inputs.

\begin{table}[t]
  \caption{Coverage of input properties}
    \label{tab:inputpropertycoverage}
    \begin{tabularx}{\linewidth}{ |X | r| r| r|}
      \hline
      
      \textbf{Approach} & \textbf{\% input properties covered} \\
      \hline
            \texttt{DocTer} & 15\% \\
      \texttt{DeepRel} & 16\% \\

      \tool{} & \textbf{31\%} \\     
\hline
  \end{tabularx}
\end{table}

\textbf{Input diversity.} Table \ref{tab:inputpropertycoverage} shows the experimental results. 
The inputs used by \tool{} in its test cases cover two times more input properties 
than the inputs used in the test cases generated by \texttt{DeepRel} and \texttt{DocTer}. 
While the inputs selected by \texttt{DeepRel} and \texttt{DocTer} cover  only 16\% and 15\% of the possible input properties,
\tool{} achieves an average of 37\% property coverage. 
This suggests that the diversity of the inputs contributed to the stronger performance of \tool{}.

\begin{table}[t]
      \caption{The distribution of outcomes running the test cases produced by \texttt{DeepRel}, \tool{}, and \texttt{DocTer}. \tool{} produces test cases with more diverse outcomes, indicating that it is better at uncovering corner cases. IAE refers to \texttt{InvalidArgumentError}}
    \label{tab:exceptions}
    \begin{tabular}{ |l | r| r| r| }
      \hline
      
      \textbf{Exception type} & \textbf{DeepRel}  & \textbf{DocTer} & \textbf{\tool{}} \\
      \hline
      None (no errors) & 77\%  & 13\%  & 24\%  \\
      IAE &  15\% & \textless 1\% &  2\%  \\
      \texttt{ValueError} & 6\% &41\%  &   7\% \\
      \texttt{TypeError} & 1\% &  46\% &  19\%  \\ 
      Other errors &  \textless 1\%  & \textless 1\%  & 48\% \\
      \hline
  \end{tabular}
\end{table}

\textbf{Output diversity.} We further investigate if the increased input diversity leads to more diverse library behaviors.
To do so,
we analyze the result of each generated test case by investigating the number of occurrences of each type of input constraint that was not satisfied.
We count the number of times each type of exception is thrown.
Note that we do not consider crashes among these outcomes (crashes are rare occurrences leading to the termination of the running process).

The proportion of each result type (e.g., a successful run without errors, or a particular exception type) is shown in Table \ref{tab:exceptions}.
\tool{} achieves a distribution with diverse test outcomes  while \texttt{DeepRel} has a greater proportion of successful executions of TensorFlow.
We stress that a high proportion of valid inputs is usually desirable 
but maybe achieved at the cost of failing to explore uncovered behaviors, for example, if a fuzzer uses the same valid input over and over again.
Our experimental result suggests that
\tool{} achieves a high diversity of outcomes;
\tool{} triggers up to 30 types of exceptions, ranging from \texttt{RecursionOverflow}, \texttt{UnicodeDecodeError}, and \texttt{ResourceExhaustedError} (categorized as ``Other errors'' in Table \ref{tab:exceptions}),   
while \texttt{DeepRel} triggers only 7 types of exceptions.
\texttt{DocTer} only triggers 4 types of exceptions, with the vast majority of them \texttt{InvalidArgumentError} and \texttt{ValueError}.
The experimental results validate our finding that \tool{} successfully triggers a greater number of different behaviors (and corner cases) compared to \texttt{DeepRel} and \texttt{DocTer}.

\subsubsection{RQ3. Generating valid inputs}

\tool{} performs input constraint inference. 
We study if the inferred input constraints are 
precise enough for producing inputs  satisfying the actual input constraints.

\texttt{DocTer} generates just valid inputs 13\% of the time, underperforming \tool{} which produces valid inputs 24\% of the time. 
This validates our initial intuition for using active learning. 
The better performance of \tool{} in generating valid inputs indicates that active learning may be more successful in inferring input constraints than \texttt{DocTer}'s use of the API documentation, which may be incomplete~\cite{xie2022docter}.

\texttt{DeepRel} uses \texttt{FreeFuzz} as its test generator, and therefore, will produce the same output as FreeFuzz.
Their input generation strategy is to mutate the seed inputs.    
As seed inputs are always semantically valid inputs, the vast majority of inputs generated by \texttt{FreeFuzz} and \texttt{DeepRel} are valid.
However, as discussed in the previous section, 
a large proportion of valid inputs may imply that the fuzzer is using similar inputs repeatedly, leading to redundancies.
Indeed, based on Table \ref{tab:exceptions}, \texttt{DeepRel} has a lower diversity of inputs, which may have led to a lower chance of generating crashing inputs (Table \ref{tab:new_vulns}).

\begin{table}[t]
  \caption{Coverage of the functions in the API. We consider an API function covered if the tool generates valid inputs. The numbers in parenthesis indicate the proportion of the API that \tool{} accepts the hypothesized input constraints for.}
    \label{tab:coverage}
    \begin{tabular}{ |l | r| r| }
      \hline
      
      \textbf{Approach} & \textbf{API Coverage} & \textbf{\# of functions covered}  \\
      \hline

      \texttt{DocTer} & 12\% & 956 \\

      \texttt{DeepRel} & 30\% & 1902  \\
      \tool{} & 37\% & 2362 \\
\hline
  \end{tabular}
\end{table}

\begin{table}[t]
  \caption{Proportion of valid inputs generated}
    \label{tab:fpr}
    \begin{tabular}{ |l | r|}
      \hline
      \textbf{Approach} & \textbf{\% of valid test cases}  \\
      \hline
      Random selection of inputs & 1\% \\
            \texttt{DocTer}   & 13\% \\
      \tool{}  & 24\%  \\
      \texttt{DeepRel} & 77\% \\
      \tool{} (valid input mode) & \textbf{80\%} \\
      
\hline
  \end{tabular}
\end{table}

\textbf{API Coverage.}
Table \ref{tab:coverage} shows the API Coverage obtained by \tool{} and the baseline tools.
\tool{} successfully invokes 37\% of the functions in the API.
In contrast, the strongest baseline, \texttt{DeepRel}, generates valid inputs for 30\% of the API functions.
The results indicate that \tool{} is able to generate valid inputs for a greater proportion of the API than existing techniques.

Table \ref{tab:fpr} shows the proportion of valid inputs generated. 
We compare \tool{} against \texttt{DocTer} as well 
as a simple baseline that randomly selects inputs used in the libraries' test suite.
While \tool{} generates  valid inputs 24\% of the time considering all three input generation modes,
\tool{} produces valid inputs 80\%  of the time in its valid input generation mode (after inferring the input constraints).
This is higher than the proportion of valid inputs generated by both \texttt{DocTer} and \texttt{DeepRel}.
Overall, this demonstrates the benefit of the active selection approach for input constraint inference.

\subsubsection{RQ4. Ablation analysis} For a deeper analysis, we perform an ablation study on \tool{}.
\tool{}$^{-}$ refers to a version of \tool{} where
inputs are sampled from the input categories, but there is no active learner posing queries and no inference of the input constraints (removing \circled{2} in Figure \ref{fig:overview}).
\tool{}$^{--}$ refer to a version of \tool{} where 
inputs are selected randomly (removing both \circled{1} and \circled{2} in Figure \ref{fig:overview}).

\begin{table}[t]
  \caption{Ablation analysis of the components in \tool{}. \% valid is the proportion of valid inputs that are generated. \tool{}$^{-}$ removes active learning. \tool{}$^{--}$ removes the use of input properties and active learning.}
    \label{tab:ablation}
    \begin{tabular}{ |l | r| r| r|}
      \hline
      
      \textbf{Approach} & \textbf{Property } & \textbf{\% valid} &  \textbf{\# crashes} \\
      & \textbf{coverage} & & \\
      \hline
      
      \tool{} & 31\% & \textbf{24\%} & \textbf{168} \\
      \tool{}$^{-}$ & \textbf{93\%} & 1\% &  112
       \\ 
    
      \tool{}$^{--}$ & 84\% & 1\% & 52 
      \\

\hline
  \end{tabular}
\end{table}

Table \ref{tab:ablation} shows the experimental results of the ablation analysis.
Without using active learning to infer input constraints, the number of crashes 
found by \tool{}$^{-}$ drops from 168 to 122, a 26\% decline.
Without using active learning, \tool{}$^{-}$ does not drive the test executor toward valid inputs.
While it is able to cover a higher proportion of properties (93\%), the majority of the inputs (99\%) are invalid.

Without the input properties, \tool{}$^{--}$ selects inputs entirely at random.
The number of detected crashes 
substantially drops to just 52, a third of the original number of crashing inputs found.
The majority of inputs selected are invalid; only 1\% of them are valid.
It spends most of its test budget using inputs that invokes the library with errors.

The experimental results indicates that higher input diversity alone is not enough. 
Having a valid input proportion that is too low hinders the ability to find crashing inputs.
Overall, our experimental results suggest that the input properties are essential to \tool{} and that  
active learning  substantially boosts the effectiveness of \tool{}.

\section{Discussion and Limitations}
\label{sec:discussion}

Our experiments demonstrate that \tool{} outperforms the existing fuzzers in generating crashing inputs
to TensorFlow and PyTorch.
Our analysis
suggests that \tool{} is effective due to the combination of both the higher diversity of inputs and the higher proportion of valid inputs.
These improvements stem from the effectiveness of active learning in input constraint inference.

\textbf{Effectiveness of active learning.} Active learning is effective in our task as we encoded the domain knowledge of deep learning libraries in the input properties.
This allows \tool{} to successfully infer the input constraints.
Had the input properties not correctly encoded the input constraints, a hypothesis would not express meaningful properties.
Once \tool{} infers the input constraints, the majority of inputs generated are valid. 
This is an improvement compared to the prior approach of extracting constraints from documentation.

\textbf{Limitations.} Next, we discuss some limitations of \tool{}.
The active learner poses queries that are answered through the invocation of the library.
This is a form of dynamic program analysis.
it, therefore, inherits the limitations of dynamic analysis; the observed behaviors are an underapproximation of the actual behaviors of a program.
Consequently, the model of the input constraints hypothesized by \tool{} may not capture some properties of the true input constraints of the library.
We leave the investigation of other methods of input constraint inference for future work.

\section{Related Work}
\label{sec:related}

\textbf{Fuzzing deep learning models and systems.} Researchers have proposed approaches to assess the security of deep learning systems.
Existing approaches fuzzes either deep learning models~\cite{xie2019deephunter,guo2018dlfuzz,du2019deepstellar} or larger systems that use deep learning~\cite{xie2011testing,he2021testing,dwarakanath2018identifying,asyrofi2021biasfinder,soremekun2022astraea,zhou2020deepbillboard}.
Other approaches use static analysis~\cite{lagouvardos2020static,liu2021detecting}.
Some studies reveal that software deploying deep learning does not secure their models well; the weights of models can be stolen by querying the model repeatedly~\cite{juuti2019prada,jagielski2020high}.
\tool{} fuzzes deep learning libraries rather than individual models or systems that use deep learning.

\textbf{Fuzzing deep learning libraries.} Several approaches have been 
proposed for testing deep learning libraries. 
Several approaches detect bugs through  metamorphic and differential testing~\cite{pham2019cradle,wang2022eagle,wang2020deep,guo2020audee,xie2019deephunter}.
These approaches check for different behaviors when the same behavior is expected, e.g. a similar function invoked with the same inputs on TensorFlow and PyTorch.
Another approach targets precision errors in TensorFlow~\cite{zhang2021predoo}.
Crucially, these previous studies overlook the systematic selection of inputs for minimizing redundancy. 

ExAIS~\cite{schumi2022exais} uses 
specifications of the deep learning layers for fuzzing.
As it requires expert analysis and manual writing of specifications, its scalability is limited.
The closest approaches to \tool{} are \texttt{DocTer}~\cite{xie2022docter} and \texttt{DeepRel}~\cite{deng2022fuzzing}, which have been discussed and used in our experiments.

\textbf{Fuzzing other libraries.} Recent research has also proposed to fuzz libraries.
Some approaches aim to generating valid inputs for libraries in specific languages, e.g. Rust~\cite{jiang2021rulf,takashima2021syrust}.
Some studies propose approaches for constructing fuzz drivers~\cite{babic2019fudge,ispoglou2020fuzzgen,zhang2021apicraft}, 
e.g. library calls to prepare the complex inputs required to invoke the library.
\tool{} has a similar goal of 
generating valid inputs
but does so through active learning to infer the input constraints.

\textbf{Selecting inputs.}
Several studies~\cite{rebert2014optimizing,wang2021syzvegas,pailoor2018moonshine,zong2020fuzzguard,chen2019enfuzz} propose methods of selecting good inputs for fuzzing.
Some methods optimize for code coverage~\cite{rebert2014optimizing,wang2021syzvegas,pailoor2018moonshine,bohme2016coverage} or filtering out inputs predicted not to reach a target program location~\cite{zong2020fuzzguard}.
Unlike these approaches, \tool{} selects inputs that may glean more information about the input constraints.

\textbf{Input validation.} 
\tool{} addresses the problem of generating inputs that pass input validation checks through input constraint inference.
Several approaches~\cite{corina2017difuze,liu2020fans,he2021sofi,peng2018t,wang2010taintscope} use static analysis  to address the problem.
DriFuzz~\cite{shen2022drifuzz} 
proposes a method of 
generating high-quality initial seed inputs.
Different from these approaches, \tool{} uses active learning to learn the input constraints to generate valid inputs.

\textbf{Active Learning.}
Our approach uses active learning~\cite{cambronero2019active,settles2009active,angluin1987learning,angluin1988queries}, which queries an oracle and learns from its feedback.
In classification tasks, active learning is used to query for labels of informative data instance when labeling every instance is too difficult~\cite{settles2009active}.
Recent work uses active learning to learn models of programs, and then regenerate programs using the models to remove undesired behaviors~\cite{vasilakis2021supply,shen2021active}.
\tool{} uses active learning to learn models of input constraints.

\section{Conclusion}
\label{sec:conclusion}

In this study, we address the problem of generating  crashing inputs to deep learning libraries.
Our approach, \tool{}, uses active learning to infer the input constraints of the libraries' API during fuzzing.
\tool{} has two advantages over existing approaches.
Firstly, \tool{} infers the input constraints without the use of documented specifications.
Secondly, its use of active learning guides the selection of a diverse set of inputs during fuzzing. 
These advantages address the challenge of generating semantically-valid inputs as well as the challenge of reducing input redundancy, which is only partially addressed and overlooked by the previous studies, respectively.
Our experiments show that addressing both challenges is crucial.
13 CVEs have been assigned to vulnerabilities found by \tool{}.
The source code of \tool{} can be found at \url{https://github.com/skipfuzz/skipfuzz}.

\balance
\bibliographystyle{plain}

\bibliography{skipfuzz.bib}

\begin{thebibliography}{10}

\bibitem{ossfuzz}
{OSS-Fuzz}.
\newblock https://github.com/google/oss-fuzz, 2022.
\newblock Accessed: 2022-10-10.

\bibitem{replication}
{SkipFuzz}'s {GitHub} repository.
\newblock https://github.com/skipfuzz/skipfuzz, 2022.

\bibitem{tfsec}
{TensorFlow} security policy.
\newblock \url{https://github.com/tensorflow/tensorflow/security/policy}, 2022.
\newblock Accessed: 2022-04-20.

\bibitem{angluin1987learning}
Dana Angluin.
\newblock Learning regular sets from queries and counterexamples.
\newblock {\em Information and computation}, 75(2):87--106, 1987.

\bibitem{angluin1988queries}
Dana Angluin.
\newblock Queries and concept learning.
\newblock {\em Machine learning}, 2(4):319--342, 1988.

\bibitem{asyrofi2021biasfinder}
Muhammad~Hilmi Asyrofi, Zhou Yang, Imam Nur~Bani Yusuf, Hong~Jin Kang, Ferdian
  Thung, and David Lo.
\newblock Biasfinder: Metamorphic test generation to uncover bias for sentiment
  analysis systems.
\newblock {\em IEEE Transactions on Software Engineering (TSE)}, 2021.

\bibitem{babic2019fudge}
Domagoj Babi{\'c}, Stefan Bucur, Yaohui Chen, Franjo Ivan{\v{c}}i{\'c}, Tim
  King, Markus Kusano, Caroline Lemieux, L{\'a}szl{\'o} Szekeres, and Wei Wang.
\newblock Fudge: fuzz driver generation at scale.
\newblock In {\em Proceedings of the 2019 27th ACM Joint Meeting on European
  Software Engineering Conference and Symposium on the Foundations of Software
  Engineering (ESEC/FSE 2019)}, pages 975--985, 2019.

\bibitem{bohme2016coverage}
Marcel B{\"o}hme, Van-Thuan Pham, and Abhik Roychoudhury.
\newblock Coverage-based greybox fuzzing as markov chain.
\newblock In {\em Proceedings of the 2016 ACM SIGSAC Conference on Computer and
  Communications Security}, pages 1032--1043, 2016.

\bibitem{cambronero2019active}
Jos{\'e}~P Cambronero, Thurston~HY Dang, Nikos Vasilakis, Jiasi Shen, Jerry Wu,
  and Martin~C Rinard.
\newblock Active learning for software engineering.
\newblock In {\em Proceedings of the 2019 ACM SIGPLAN International Symposium
  on New Ideas, New Paradigms, and Reflections on Programming and Software},
  pages 62--78, 2019.

\bibitem{chen2019enfuzz}
Yuanliang Chen, Yu~Jiang, Fuchen Ma, Jie Liang, Mingzhe Wang, Chijin Zhou, Xun
  Jiao, and Zhuo Su.
\newblock {EnFuzz}: Ensemble fuzzing with seed synchronization among diverse
  fuzzers.
\newblock In {\em 28th USENIX Security Symposium (USENIX Security 19)}, pages
  1967--1983, 2019.

\bibitem{corina2017difuze}
Jake Corina, Aravind Machiry, Christopher Salls, Yan Shoshitaishvili, Shuang
  Hao, Christopher Kruegel, and Giovanni Vigna.
\newblock Difuze: Interface aware fuzzing for kernel drivers.
\newblock In {\em Proceedings of the 2017 ACM SIGSAC Conference on Computer and
  Communications Security (CCS 2017)}, pages 2123--2138, 2017.

\bibitem{deng2022fuzzing}
Yinlin Deng, Chenyuan Yang, Anjiang Wei, and Lingming Zhang.
\newblock Fuzzing deep-learning libraries via automated relational {API}
  inference.
\newblock In {\em 2022 ACM Joint European Software Engineering Conference and
  Symposium on the Foundations of Software Engineering (ESEC/FSE 2022)}, 2022.

\bibitem{du2019deepstellar}
Xiaoning Du, Xiaofei Xie, Yi~Li, Lei Ma, Yang Liu, and Jianjun Zhao.
\newblock {DeepStellar}: Model-based quantitative analysis of stateful deep
  learning systems.
\newblock In {\em Proceedings of the 2019 27th ACM Joint Meeting on European
  Software Engineering Conference and Symposium on the Foundations of Software
  Engineering (ESEC/FSE 2019)}, pages 477--487, 2019.

\bibitem{dwarakanath2018identifying}
Anurag Dwarakanath, Manish Ahuja, Samarth Sikand, Raghotham~M Rao,
  RP~Jagadeesh~Chandra Bose, Neville Dubash, and Sanjay Podder.
\newblock Identifying implementation bugs in machine learning based image
  classifiers using metamorphic testing.
\newblock In {\em Proceedings of the 27th ACM SIGSOFT International Symposium
  on Software Testing and Analysis (ISSTA 2018)}, pages 118--128, 2018.

\bibitem{gebser2008user}
Martin Gebser, Roland Kaminski, Benjamin Kaufmann, Max Ostrowski, Torsten
  Schaub, and Sven Thiele.
\newblock A user’s guide to gringo, clasp, clingo, and iclingo.
\newblock 2008.

\bibitem{guo2018dlfuzz}
Jianmin Guo, Yu~Jiang, Yue Zhao, Quan Chen, and Jiaguang Sun.
\newblock {DLFuzz}: Differential fuzzing testing of deep learning systems.
\newblock In {\em Proceedings of the 2018 26th ACM Joint Meeting on European
  Software Engineering Conference and Symposium on the Foundations of Software
  Engineering (ESEC/FSE 2018)}, pages 739--743, 2018.

\bibitem{guo2020audee}
Qianyu Guo, Xiaofei Xie, Yi~Li, Xiaoyu Zhang, Yang Liu, Xiaohong Li, and Chao
  Shen.
\newblock Audee: Automated testing for deep learning frameworks.
\newblock In {\em 2020 35th IEEE/ACM International Conference on Automated
  Software Engineering (ASE 2020)}, pages 486--498. IEEE, 2020.

\bibitem{he2021testing}
Pinjia He, Clara Meister, and Zhendong Su.
\newblock Testing machine translation via referential transparency.
\newblock In {\em 2021 IEEE/ACM 43rd International Conference on Software
  Engineering (ICSE 2021)}, pages 410--422. IEEE, 2021.

\bibitem{he2021sofi}
Xiaoyu He, Xiaofei Xie, Yuekang Li, Jianwen Sun, Feng Li, Wei Zou, Yang Liu,
  Lei Yu, Jianhua Zhou, Wenchang Shi, et~al.
\newblock {SoFi}: Reflection-augmented fuzzing for javascript engines.
\newblock In {\em Proceedings of the 2021 ACM SIGSAC Conference on Computer and
  Communications Security (CSS 2021)}, pages 2229--2242, 2021.

\bibitem{islam2019comprehensive}
Md~Johirul Islam, Giang Nguyen, Rangeet Pan, and Hridesh Rajan.
\newblock A comprehensive study on deep learning bug characteristics.
\newblock In {\em Proceedings of the 2019 27th ACM Joint Meeting on European
  Software Engineering Conference and Symposium on the Foundations of Software
  Engineering (ESEC/FSE 2019)}, pages 510--520, 2019.

\bibitem{ispoglou2020fuzzgen}
Kyriakos Ispoglou, Daniel Austin, Vishwath Mohan, and Mathias Payer.
\newblock {FuzzGen}: Automatic fuzzer generation.
\newblock In {\em 29th USENIX Security Symposium (USENIX Security 20)}, pages
  2271--2287, 2020.

\bibitem{jagielski2020high}
Matthew Jagielski, Nicholas Carlini, David Berthelot, Alex Kurakin, and Nicolas
  Papernot.
\newblock High accuracy and high fidelity extraction of neural networks.
\newblock In {\em 29th USENIX security symposium (USENIX Security 20)}, pages
  1345--1362, 2020.

\bibitem{jia2020empirical}
Li~Jia, Hao Zhong, Xiaoyin Wang, Linpeng Huang, and Xuansheng Lu.
\newblock An empirical study on bugs inside {TensorFlow}.
\newblock In {\em International Conference on Database Systems for Advanced
  Applications}, pages 604--620. Springer, 2020.

\bibitem{jia2021symptoms}
Li~Jia, Hao Zhong, Xiaoyin Wang, Linpeng Huang, and Xuansheng Lu.
\newblock The symptoms, causes, and repairs of bugs inside a deep learning
  library.
\newblock {\em Journal of Systems and Software (JSS)}, 177:110935, 2021.

\bibitem{jiang2021rulf}
Jianfeng Jiang, Hui Xu, and Yangfan Zhou.
\newblock {RULF}: Rust library fuzzing via api dependency graph traversal.
\newblock In {\em 2021 36th IEEE/ACM International Conference on Automated
  Software Engineering (ASE 2021)}, pages 581--592. IEEE, 2021.

\bibitem{juuti2019prada}
Mika Juuti, Sebastian Szyller, Samuel Marchal, and N~Asokan.
\newblock Prada: protecting against dnn model stealing attacks.
\newblock In {\em 2019 IEEE European Symposium on Security and Privacy
  (EuroS\&P)}, pages 512--527. IEEE, 2019.

\bibitem{lagouvardos2020static}
Sifis Lagouvardos, Julian Dolby, Neville Grech, Anastasios Antoniadis, and
  Yannis Smaragdakis.
\newblock Static analysis of shape in {TensorFlow} programs.
\newblock In {\em 34th European Conference on Object-Oriented Programming
  (ECOOP 2020)}. Schloss Dagstuhl-Leibniz-Zentrum f{\"u}r Informatik, 2020.

\bibitem{liu2020fans}
Baozheng Liu, Chao Zhang, Guang Gong, Yishun Zeng, Haifeng Ruan, and Jianwei
  Zhuge.
\newblock {FANS}: Fuzzing {Android} native system services via automated
  interface analysis.
\newblock In {\em 29th USENIX Security Symposium (USENIX Security 20)}, pages
  307--323, 2020.

\bibitem{liu2021detecting}
Chen Liu, Jie Lu, Guangwei Li, Ting Yuan, Lian Li, Feng Tan, Jun Yang, Liang
  You, and Jingling Xue.
\newblock Detecting {TensorFlow} program bugs in real-world industrial
  environment.
\newblock In {\em 2021 36th IEEE/ACM International Conference on Automated
  Software Engineering (ASE 2021)}, pages 55--66. IEEE, 2021.

\bibitem{miller2020relevance}
Barton Miller, Mengxiao Zhang, and Elisa Heymann.
\newblock The relevance of classic fuzz testing: Have we solved this one?
\newblock {\em IEEE Transactions on Software Engineering (TSE)}, 2020.

\bibitem{miller1990empirical}
Barton~P Miller, Louis Fredriksen, and Bryan So.
\newblock An empirical study of the reliability of {UNIX} utilities.
\newblock {\em Communications of the ACM}, 33(12):32--44, 1990.

\bibitem{padhye2019semantic}
Rohan Padhye, Caroline Lemieux, Koushik Sen, Mike Papadakis, and Yves Le~Traon.
\newblock Semantic fuzzing with {Zest}.
\newblock In {\em Proceedings of the 28th ACM SIGSOFT International Symposium
  on Software Testing and Analysis (ISSTA 2019)}, pages 329--340, 2019.

\bibitem{pailoor2018moonshine}
Shankara Pailoor, Andrew Aday, and Suman Jana.
\newblock {MoonShine}: Optimizing {OS} fuzzer seed selection with trace
  distillation.
\newblock In {\em 27th USENIX Security Symposium (USENIX Security 18)}, pages
  729--743, 2018.

\bibitem{peng2018t}
Hui Peng, Yan Shoshitaishvili, and Mathias Payer.
\newblock T-fuzz: fuzzing by program transformation.
\newblock In {\em 2018 IEEE Symposium on Security and Privacy (S\&P)}, pages
  697--710. IEEE, 2018.

\bibitem{pham2019cradle}
Hung~Viet Pham, Thibaud Lutellier, Weizhen Qi, and Lin Tan.
\newblock {CRADLE}: cross-backend validation to detect and localize bugs in
  deep learning libraries.
\newblock In {\em 2019 IEEE/ACM 41st International Conference on Software
  Engineering (ICSE 2019)}, pages 1027--1038. IEEE, 2019.

\bibitem{rebert2014optimizing}
Alexandre Rebert, Sang~Kil Cha, Thanassis Avgerinos, Jonathan Foote, David
  Warren, Gustavo Grieco, and David Brumley.
\newblock Optimizing seed selection for fuzzing.
\newblock In {\em 23rd USENIX Security Symposium (USENIX Security 14)}, pages
  861--875, 2014.

\bibitem{sato2021dirty}
Takami Sato, Junjie Shen, Ningfei Wang, Yunhan Jia, Xue Lin, and Qi~Alfred
  Chen.
\newblock Dirty road can attack: Security of deep learning based automated lane
  centering under $\{$Physical-World$\}$ attack.
\newblock In {\em 30th USENIX Security Symposium (USENIX Security 21)}, pages
  3309--3326, 2021.

\bibitem{schumi2022exais}
Richard Schumi and Jun Sun.
\newblock {ExAIS}: Executable ai semantics.
\newblock In {\em 2022 IEEE/ACM 41st International Conference on Software
  Engineering (ICSE 2022)}, 2022.

\bibitem{settles2009active}
Burr Settles.
\newblock Active learning literature survey.
\newblock 2009.

\bibitem{shen2021active}
Jiasi Shen and Martin~C Rinard.
\newblock Active learning for inference and regeneration of applications that
  access databases.
\newblock {\em ACM Transactions on Programming Languages and Systems (TOPLAS)},
  42(4):1--119, 2021.

\bibitem{shen2022drifuzz}
Zekun Shen, Ritik Roongta, and Brendan Dolan-Gavitt.
\newblock Drifuzz: Harvesting bugs in device drivers from golden seeds.
\newblock In {\em 31st USENIX Security Symposium (USENIX Security 22)}, pages
  1275--1290, 2022.

\bibitem{soremekun2022astraea}
Ezekiel Soremekun, Sakshi~Sunil Udeshi, and Sudipta Chattopadhyay.
\newblock Astraea: Grammar-based fairness testing.
\newblock {\em IEEE Transactions on Software Engineering (TSE)}, 2022.

\bibitem{takashima2021syrust}
Yoshiki Takashima, Ruben Martins, Limin Jia, and Corina~S
  P{\u{a}}s{\u{a}}reanu.
\newblock {SyRust}: automatic testing of {Rust} libraries with semantic-aware
  program synthesis.
\newblock In {\em Proceedings of the 42nd ACM SIGPLAN International Conference
  on Programming Language Design and Implementation (PLDI 2021)}, pages
  899--913, 2021.

\bibitem{vasilakis2021supply}
Nikos Vasilakis, Achilles Benetopoulos, Shivam Handa, Alizee Schoen, Jiasi
  Shen, and Martin~C Rinard.
\newblock Supply-chain vulnerability elimination via active learning and
  regeneration.
\newblock In {\em Proceedings of the 2021 ACM SIGSAC Conference on Computer and
  Communications Security (CCS 2021)}, pages 1755--1770, 2021.

\bibitem{wang2021syzvegas}
Daimeng Wang, Zheng Zhang, Hang Zhang, Zhiyun Qian, Srikanth~V Krishnamurthy,
  and Nael Abu-Ghazaleh.
\newblock {SyzVegas}: Beating kernel fuzzing odds with reinforcement learning.
\newblock In {\em 30th USENIX Security Symposium (USENIX Security 21)}, pages
  2741--2758, 2021.

\bibitem{wang2022eagle}
Jiannan Wang, Thibaud Lutellier, Shangshu Qian, Hung~Viet Pham, and Lin Tan.
\newblock Eagle: Creating equivalent graphs to test deep learning libraries.
\newblock In {\em 2022 IEEE/ACM 41st International Conference on Software
  Engineering (ICSE 2022)}, 2022.

\bibitem{wang2010taintscope}
Tielei Wang, Tao Wei, Guofei Gu, and Wei Zou.
\newblock Taintscope: A checksum-aware directed fuzzing tool for automatic
  software vulnerability detection.
\newblock In {\em 2010 IEEE Symposium on Security and Privacy (S\&P}, pages
  497--512. IEEE, 2010.

\bibitem{wang2020deep}
Zan Wang, Ming Yan, Junjie Chen, Shuang Liu, and Dongdi Zhang.
\newblock Deep learning library testing via effective model generation.
\newblock In {\em Proceedings of the 28th ACM Joint Meeting on European
  Software Engineering Conference and Symposium on the Foundations of Software
  Engineering (ESEC/FSE 2020)}, pages 788--799, 2020.

\bibitem{wei2022free}
Anjiang Wei, Yinlin Deng, Chenyuan Yang, and Lingming Zhang.
\newblock Free lunch for testing: Fuzzing deep-learning libraries from open
  source.
\newblock In {\em 2022 IEEE/ACM 41st International Conference on Software
  Engineering (ICSE 2022)}, 2022.

\bibitem{xie2022docter}
Danning Xie, Yitong Li, Mijung Kim, Hung~Viet Pham, Lin Tan, Xiangyu Zhang, and
  Michael~W Godfrey.
\newblock {DocTer}: documentation-guided fuzzing for testing deep learning api
  functions.
\newblock In {\em Proceedings of the 31st ACM SIGSOFT International Symposium
  on Software Testing and Analysis (ISSTA 2022)}, pages 176--188, 2022.

\bibitem{xie2019deephunter}
Xiaofei Xie, Lei Ma, Felix Juefei-Xu, Minhui Xue, Hongxu Chen, Yang Liu,
  Jianjun Zhao, Bo~Li, Jianxiong Yin, and Simon See.
\newblock {DeepHunter}: a coverage-guided fuzz testing framework for deep
  neural networks.
\newblock In {\em Proceedings of the 28th ACM SIGSOFT International Symposium
  on Software Testing and Analysis (ISSTA 2019)}, pages 146--157, 2019.

\bibitem{xie2011testing}
Xiaoyuan Xie, Joshua~WK Ho, Christian Murphy, Gail Kaiser, Baowen Xu, and
  Tsong~Yueh Chen.
\newblock Testing and validating machine learning classifiers by metamorphic
  testing.
\newblock {\em Journal of Systems and Software (JSS)}, 84(4):544--558, 2011.

\bibitem{xintan2022deeplearningsupplychain}
Minghui Zhou Li~Zhang Xin~Tan, Kai~Gao.
\newblock An exploratory study of deep learning supply chain.
\newblock In {\em 2022 IEEE/ACM 41st International Conference on Software
  Engineering (ICSE)}, 2022.

\bibitem{zhang2021apicraft}
Cen Zhang, Xingwei Lin, Yuekang Li, Yinxing Xue, Jundong Xie, Hongxu Chen,
  Xinlei Ying, Jiashui Wang, and Yang Liu.
\newblock {APICraft}: Fuzz driver generation for closed-source {SDK} libraries.
\newblock In {\em 30th USENIX Security Symposium (USENIX Security 21)}, pages
  2811--2828, 2021.

\bibitem{zhang2021predoo}
Xufan Zhang, Ning Sun, Chunrong Fang, Jiawei Liu, Jia Liu, Dong Chai, Jiang
  Wang, and Zhenyu Chen.
\newblock Predoo: precision testing of deep learning operators.
\newblock In {\em Proceedings of the 30th ACM SIGSOFT International Symposium
  on Software Testing and Analysis (ISSTA 2021)}, pages 400--412, 2021.

\bibitem{zhou2020deepbillboard}
Husheng Zhou, Wei Li, Zelun Kong, Junfeng Guo, Yuqun Zhang, Bei Yu, Lingming
  Zhang, and Cong Liu.
\newblock Deepbillboard: Systematic physical-world testing of autonomous
  driving systems.
\newblock In {\em 2020 IEEE/ACM 42nd International Conference on Software
  Engineering (ICSE)}, pages 347--358. IEEE, 2020.

\bibitem{zong2020fuzzguard}
Peiyuan Zong, Tao Lv, Dawei Wang, Zizhuang Deng, Ruigang Liang, and Kai Chen.
\newblock {FuzzGuard}: Filtering out unreachable inputs in directed grey-box
  fuzzing through deep learning.
\newblock In {\em 29th USENIX Security Symposium (USENIX Security 20)}, pages
  2255--2269, 2020.

\end{thebibliography}

\end{document}